\input amstex
\pageno=1

\magnification=1200
\loadmsbm
\loadmsam
\loadeufm
\input amssym
\UseAMSsymbols

\TagsOnRight

\hsize172 true mm
\vsize212 true mm
\voffset=20 true mm
\hoffset=0 true mm
\baselineskip 7.0 true mm plus0.4 true mm minus0.4 true  mm

\def\P{\partial}

\def\A{\alpha}
\def\B{\beta}

\def\G{\gamma}

\def\N{\eta}
\def\D{\delta}
\def\X{\xi}
\def\F{\psi}
\def\SI{\sigma}

\def\W{\widetilde}

\vskip 0.4cm
\noindent

\centerline{\bf FORMULAS FOR $A_n$ AND $B_n$--SOLUTIONS OF WDVV
EQUATIONS}
\vskip 0.4cm
\centerline{S.Œ. NATANZON}
\vskip 0.4cm

\centerline{1. Introduction}
\vskip 0.4cm
\noindent

1. WDVV equations were appeared in works of E.Witten [W1],
R.Dijkgraaf, E.Verlinde, H.Verlinde [DVV]
as equations for primary free energy of
two-dimensional topological field theories.
The simplest non-trivial solutions of WDVV equations are
$A_n$ and $B_n$-potentials, which describe also metrics of
K.Saito on spaces of versal deformation of $A_n$ and
$B_n$-singula\-rities. According to E.Witten [W2] coefficients
of $A_n$--potentials form intersection
numbers  of Mumford--Morita--Miller classes of
a moduli space of spheres with punctures and $n$--spin structures.
The potentials are known for $n\leqslant 4$ [W2, D1, D2].
We find  these potentials for all $n$, using its connection with
Gelfand-Dikii $(n-k\P V)$ hierarchy.
In passing we give recurrence formulas for
coefficients of dispersionless KP hierarchy.
Results of J.-B.Zuber [Z]
give a reduction of $B_n$--potentials to $A_n$--potentials.

I would like to thank B.Dubrovin for fruitful discussions.

\vskip 0.4cm
\centerline{2. Main theorems}
\vskip 0.4cm

For calculation of the $A_{n-1}$-potential we use combinatorial
coefficients $P_n(i_1...i_m)$, where $n$ and $s_j$ are natural
numbers. Put us
$$P_n(i)=n,\quad P_n(s_1...s_m)=C^m_n-\sum^{m-1}_{q=1}
P_n(s_1...s_q)C^{m-q}_{n-q-(s_1+\dotsb+s_q)},$$
where here and later
$$0!=1, C^t_p=\frac{p!}{t!(p-t)!}\ \text{for}\ p\geqslant t\geqslant 0
\ \text{and} \ C^t_p=0\ \text{in another cases}.$$

Put us $x_0=0$. For $q>0$ we recurrently define polynomials
$x_{-q}$ $(x_1,...,x_{n-1})$ putting
$$x_{-q}=-\frac{1}{n}
\sum_{m=2}^{\infty}\sum P_{n}
(s_1\dots s_m)x_{n-s_1}\dotsb x_{n-s_m},$$
where the second sum is taken by all natural numbers $(s_1,...,s_m)$
such that $\sum_{i=1}^m s_i=q+n+1-m$.

This polynomial has a form $$x_{-q}=\sum_{m=1}^\infty
\sum B_{qi_1\dotsb i_m}x_{i_1}
\dotsb x_{i_m},$$ where $B_{q i_1,...,i_m}$ are constants and
the second sum is taken by numbers
$(i_1,...,i_m)$ such that $1\leqslant i_1\leqslant ...\leqslant
i_m\leqslant n-1$.

Put us $$A_{i_1\dotsb i_m}=- \frac{1}{i_1 p} B_{i_1\dotsb i_m}$$
and $p$ is the number of $j$ such that $i_j=i_1$.

Consider $$F_A=\sum_{m=3}^\infty \sum A_{i_1\dotsb i_m} 
x_{i_1}\dotsb x_{i_m},$$ 
$$F_B=\sum_{m=3}^\infty\sum A_{2i_1-1\dotsb 2i_m-1}x_{i_1}\dotsb x_{i_m},$$
where the second sums is taken by $1\leqslant i_1\leqslant ...
\leqslant i_m\leqslant n-1$.

\vskip 0.4cm
\noindent
{\bf Theorem 1.} {\sl Polynomial $F_A$ is a $A_{n-1}$--potential.
Polynomial $F_B$ is a $B_{n-1}$--potential.
This means that they satisfy the equations
$$\sum^{n-1}_{\G=1}\frac{\P^3 F}{\P x_\A \P x_\B \P x_\G}
\frac{\P^3 F}{\P x_{n-\G} \P x_{\N} \P x_\X}=
\sum^{n-1}_{\G=1}\frac{\P^3 F}{\P x_\X \P x_\B \P x_\G}
\frac{\P^3 F}{\P x_{n-\G} \P x_{\N} \P x_\A},\
(\A,\B,\eta,\X=1,...,n-1),$$

$$\frac{\P^3 F}{\P x_1 \P x_\A \P x_\B}=\D_{\A+\B,n},\quad
\sum_{j=1}^{n-1}d_i \frac{\P F}{\P x_i}=d F,$$
where $d_i=1+\frac{1-i}{n}$, $d=2+\frac 2n$ for $F=F_A$ and
$d_i=1+\frac{1-i}{n-1},\ d=2+\frac{1}{n-1}$ for $F=F_B$.}

\vskip 0.4cm
\noindent
\centerline{3. Combinatorial lemma}
\vskip 0.4cm
\noindent

For natural $s, i_1,...,i_n$ and integer not negative
$j_1,...,j_n$ defind $P_s\pmatrix i_1&\ldots&i_n\\j_1&\ldots&j_n
\endpmatrix$ by recu\-rence formulas:
$$1) P_s\pmatrix i_1&\ldots&i_n\\0&\ldots&0\endpmatrix=0; \quad
2) P_s\pmatrix i\\j\endpmatrix=C^j_s
\quad \text{for}\ j>0;$$
$$3) P_s\pmatrix i_1&\ldots&i_n\\j_1&\ldots&j_n\endpmatrix=\frac{1}{n!}
C_s^{j_1+\dotsb+j_n}\frac{(j_1+\dotsb+j_n)!}{j_1!\dotsb j_n!}-$$
$$\sum^{n-1}_{q=1}P_s
\pmatrix i_1 &\ldots&i_q\\j_1 &\ldots&j_q\endpmatrix\frac{1}{(n-q)!}
C^{j_{q+1}+\dotsb+j_n}_{s-(i_1+\dotsb+i_q+j_1+\dotsb+j_q)}
\frac{(j_{q+1}+\dotsb+j_n)!}{j_{q+1}!\dotsb j_n!}$$
for $(j_1,...,j_n)\ne (0,...,0)$.

Let $\bmatrix i_1 &\ldots&i_n\\j_1 &\ldots&j_n\endbmatrix$ be the set of
all matrices, which appear from $\pmatrix i_1 &\ldots&i_n\\j_1 &\ldots&j_n
\endpmatrix$ by permu\-tation of columns.
Let $\Vmatrix i_1 &\ldots&i_n\\j_1 &\ldots&j_n\endVmatrix$ be the number
of such matrices. Put us
$$P_s\bmatrix i_1 &\ldots&i_n\\j_1 &\ldots&j_n\endbmatrix=\sum
P_s\pmatrix a_1 &\ldots&a_n\\b_1 &\ldots&b_n\endpmatrix,$$ where the sum is
taken by all
$$\pmatrix a_1 &\ldots&a_n\\b_1 &\ldots&b_n\endpmatrix\in
\bmatrix i_1 &\ldots&i_n\\j_1 &\ldots&j_n\endbmatrix$$

\vskip 0.4cm
\noindent
{\bf Lemma 1} {\sl Let $m>0, k>0$ and $j_n\geqslant 1$
for $n\leqslant m$. Then
$$P_s\bmatrix i_{1} & \ldots &i_m & i_{m+1} &\ldots &i_{m+k}\\
j_1 &\ldots &j_m & 0 &\ldots & 0\endbmatrix=$$
$$=\cases 0, \quad \text{if $s\geqslant i_1+...+i_m+j_1+...+j_m$,}\\
\frac{1}{k!}\Vmatrix i_{m+1}&\ldots& i_{m+k}\\0&\ldots &0\endVmatrix
P_s\bmatrix i_1&\ldots&i_m\\j_1&\ldots &j_m\endbmatrix, \quad
\text{if $s< i_1+...+i_m+j_1+...+j_m.$} \endcases$$}

\vskip 0.4cm
\noindent
Proof: Prove at first the lemma for $m=1$ using induction by k.
For $m=k=1$
$$P_s\bmatrix i_1 \ i_2\\j_1 \ 0\endbmatrix=
P_s\pmatrix i_1 \ i_2\\j_1 \ 0\endpmatrix+
P_s\pmatrix i_2 \ i_1\\0 \ j_1\endpmatrix=$$
$$\frac 12 C^{j_1}_s-P_s\pmatrix i_1\\j_1\endpmatrix\cdot
C^0_{s-(i_1+j_1)}+\frac 12 C^{j_1}_s-P_s\pmatrix i_2\\0\endpmatrix
\cdot C^0_{s-i_2}=C^{j_1}_s-C^{j_1}_s C^0_{s-(i_1+j_1)}=$$
$$=\cases 0, &\text{if $s\geqslant i_1+j_1,$}\\
C^{j_1}_s=P_s\bmatrix i_1\\ j_1\endbmatrix, &\text{if
$s<i_1+j_1.$}\endcases$$
Prove the lemma for $m=1, k=N$, considering that it is proved for
$m=1, k<N$. If $s\geqslant i_1+j_1$, then
$$P_1\bmatrix i_1&i_2&\ldots & i_{k+1}\\j_1&0&\ldots & 0\endbmatrix=
\frac{1}{k!}C^{j_1}_s
\Vmatrix i_2 &\ldots & i_{k+1}\\0&\ldots &  0\endVmatrix-
P_s\pmatrix i_1\\j_1\endpmatrix\cdot C^0_{s-(i_1+j_1)}\cdot
\frac{1}{k!}\Vmatrix i_2 &\ldots & i_{k+1}\\0 &\ldots & 0\endVmatrix=0.$$
If $s<i_1+j_1$ than
$$P_1\bmatrix i_1&i_2 &\ldots & i_{k+1}\\j_1 & 0 &\ldots &  0\endbmatrix=
\frac{1}{k!}C^{j_1}_s
\Vmatrix i_2 &\ldots & i_{k+1}\\0 &\ldots & 0\endVmatrix-
A \cdot C^0_{s-(i_1+j_1)}=$$
$$\frac{1}{k!}\Vmatrix i_2 &\ldots & ,i_{k+1}\\0 &\ldots & 0\endVmatrix
P_s\bmatrix i_1\\ j_1\endbmatrix.$$ Thus the lemma is proved
for $m=1$.

Prove the lemma for $m=N$ considering that it is proved for $m<N$.
Then $$P_s\bmatrix i_1 &\ldots & i_m&i_{m+1} &\ldots & i_{m+k}\\j_1
 &\ldots & j_m &  0 &\ldots &  0\endbmatrix=$$
$$\sum_{\pmatrix \A_1 &\ldots & \A_m\\ \B_1 &\ldots & \B_m\endpmatrix\in
\bmatrix i_1 &\ldots & i_m\\j_1 &\ldots & j_m\endbmatrix}\Bigl(\frac{1}
{(m+k)!} C^{\B_1+\dotsb+\B_m}_s\frac{(\B_1+\dotsb+\B_m)!}{\B_1!\dotsb \B_m!}
C^k_{m+k}\cdot$$
$$\Vmatrix i_{m+1} &\ldots & i_{m+k}\\ 0 &\ldots & 0\endVmatrix-$$
$$-\sum_{q=1}^m P_s \pmatrix \A_1 &\ldots & \A_q\\
\B_1 &\ldots & \B_q\endpmatrix
\frac{1}{(m+k-q)!} C^{\B_{q+1}+\dotsb+\B_m}_{s-(\A_1+\dotsb
+\A_q+\B_1+\dotsb +\B_q)}
\frac{(\B_{q+1}+\dotsb+\B_m)!}{\B_{q+1}!\dotsb \B_m!}
C^k_{m+k-q}\cdot$$
$$\Vmatrix i_{m+1} &\ldots & i_{m+k}\\ 0 &\ldots & 0\endVmatrix\Bigr)=
P_s\bmatrix i_1 &\ldots & i_m\\ j_1 &\ldots & j_m\endbmatrix\frac{1}{k!}
\Vmatrix i_{m+1} &\ldots & i_{m+k}\\ 0 &\ldots &  0\endVmatrix-$$
$$P_s\bmatrix i_1 &\ldots & i_m\\ j_1 &\ldots & j_m\endbmatrix
C^0_{s-(i_1+\dotsb+i_m+j_1+\dotsb+j_m)}\frac{1}{k!}
\Vmatrix i_{m+1} &\ldots &  i_{m+k}\\ 0 &\ldots & 0\endVmatrix=$$
$$=\cases 0,\quad\text{if $s\geqslant i_1+\dotsb+i_m+j_1+\dotsb+j_m,$}\\
P_s\bmatrix i_1 &\ldots & i_m\\ j_1 &\ldots & j_m\endbmatrix
\frac{1}{k!} \Vmatrix i_{m+1} &\ldots &  i_{m+k}\\ 0 &\ldots & 0\endVmatrix
, &\text{if $s<i_1+\dotsb+i_m+j_1+\dotsb+j_m$.$\square$}\endcases$$

\vskip 0.4cm
\noindent
\centerline {4. Equations for Bacher-Akhiezer function}
\vskip 0.4cm
\noindent

Consider the KP hierarchy. This is a condition of compatibility of the
infinite system of the differential equations
$$\frac{\P\F}{\P x_n}=L_n\F,\ \tag 1$$
where $$L_n=\frac{\P^n}{\P x_1^n}+\sum^n_{i=2}B^i_n(x)
\frac{\P^{n-i}}{\P^{n-i} x_1},$$ and $\F$ is
a function of type
$$\F(x,k)=\text{exp}(\sum^\infty_{j=1}x_j k^j)(1+\sum^\infty_{i=1}
\X_i k^{-i}), $$
(here $k\in \Bbb C$ belong to some neighbourhood of $\infty$ and
$x=(x_1, x_2,...)$ --- is a  finite sequence).

Put us $$\P_i=\frac{\P}{\P x_i},\ \P=\P_1.$$ A direct
calculation gives

\vskip 0.4cm
\noindent
{\bf Lemma 2.} {\sl Conditions of compatibility of (1) are
$$B^t_s=-\sum^{t-1}_{i=1} C^i_s \P^i\X_{t-i}
-\sum^{t-1}_{j=2}B^j_s\sum^{t-j-1}_{i=0}C^i_{s-j} \P^i\X_{t-i-j}
, \tag 2$$
$$\P_n\X_i=\sum^{n+i-1}_{j=1} C^j_n \P^j\X_{i+n-j}+
\sum^{n}_{k=2}B^k_n \sum^{n-k}_{j=0}C^j_{n-k} \P^j\X_{i+n-j-k}
.\quad \square\tag 3$$ }

In this case $\F$ is called a {\sl Bacher-Akhiezer function}.

Consider now the function $$\ln\F(x,k)=\sum^\infty_{j=1} x_j k^j+
\sum^\infty_{j=1}\eta_j k^{-j},$$ where
$$\X_j=\sum^\infty_{n=1}\frac {1}{n!}\sum_{i_1+\dotsb+i_n=j}
\eta_{i_1}\dotsb \eta_{i_n}. $$

\vskip 0.4cm
\noindent
{\bf Lemma 3.} {\sl Let $2\leqslant t\leqslant s$.
Then $$B^t_s=-\sum_{n=1}^\infty \sum P_s\pmatrix i_1 &\ldots & i_n\\
j_1 &\ldots & j_n\endpmatrix \P^{j_1}\eta_{i_1}\dotsb \P^{i_n}\eta_{i_n},$$
where the second sum is taken by all matrices $\pmatrix i_1 &\ldots & i_n\\
j_1 &\ldots & j_n\endpmatrix$ such that $i_m\geqslant 1, j_m\geqslant 1$ and
$i_1+\dotsb +i_n+j_1+\dotsb +j_n=t$.}

\vskip 0.4cm
\noindent
Proof: An induction by $t$. For $t=2$ according to (2),
$B^2_s=-s\P\X_1=-P_s {1\choose 1}\P\eta_1$. Prove the lemma for
$t=N$ considering that it is proved for $t<N$. According to (2)
$$B^t_s=-\sum^{t-1}_{i=1} C^i_s \P^i
\Bigr(\sum^{\infty}_{n=1} \frac{1}{n!}\sum_{i_1+\dotsb +i_n=t-i}
\eta_{i_1}\dotsb\eta_{i_n}\Bigl)+$$
$$+\sum_{j=2}^{t-1}\Bigl(\sum^\infty_{n=1}\sum_{i_1+\dotsb +j_n=j} P_s
{\pmatrix i_1 &\ldots & i_n\\j_1 &\ldots & j_n\endpmatrix}
\P^{j_1}\eta_{i_1}\dotsb \P^{j_n}\eta_{i_n}\Bigr)\cdot$$
$$\Bigl(\sum_{i=0}^{t-j-1}C^i_{s-j}\P^j
\Bigl(\sum^\infty_{n=1}\frac {1}{n!}
\sum_{i_1+\dotsb +i_n=t-i-j}\eta_{i_1}\dotsb\eta_{i_n}\Bigr)\Bigr)=$$
$$=-\sum^\infty_{n=1}\sum\Bigl({1\over{n!}} C_s^{j_1+\dotsb +j_n}
\frac{(j_1+\dotsb +j_n)!}{j_1!\dotsb j_n!}-$$
$$\sum^{n-1}_{q=1}P_s{\pmatrix i_1 &\ldots & i_q\\
j_1 &\ldots & j_q\endpmatrix}
\frac{1}{(n-q)!} C^{j_{q+1}\dotsb j_n}_{s-(i_1+\dotsb i_q+j_1+\dotsb
+j_q)}\frac {(j_{q+1}+\dotsb j_n)!}{j_{q+1}!\dotsb j_n!}\Bigr)
\P^{j_1}\eta_{i_1}\dotsb \P^{j_n}\eta_{i_n}=$$
$$=-\sum^\infty_{n=1}\sum P_s{\pmatrix i_1 &\ldots & i_n\\
j_1 &\ldots & j_n\endpmatrix}
\P^{j_1}\eta_{i_1}\dotsb \P^{j_n}\eta_{i_n},$$
where the second sums are taken by all matrices $\pmatrix i_1 &\ldots & i_n\\
j_1 &\ldots & j_n\endpmatrix$ such that $i_1+\dotsb +i_n+j_1+\dotsb +
j_n=t$, $i_m\geqslant 1, j_m\geqslant 0$.
According to lemma 1 it is possible to consider that in the last sum
$j_m>0$ for all $m$. $\square$

\vskip 0.4cm
\noindent
{\bf Lemma 4.} {\sl
$$\P_s\eta_r=\sum_{n=1}^\infty \sum P_s\pmatrix i_1 &\ldots & i_n\\
j_1 &\ldots & j_n\endpmatrix \P^{j_1}\eta_{i_1}\dotsb \P^{j_n}\eta_{i_n},$$
where the second sum is taken by all matrices $\pmatrix i_1 &\ldots & i_n\\
j_1 &\ldots & j_n\endpmatrix$ such that $i_m\geqslant 1, j_m\geqslant 1$ and
$i_1+\dotsb +i_n+j_1+\dotsb +j_n=r+s$.}

\vskip 0.4cm
\noindent
Proof: An induction by $r$. According to (3) and lemma 3 for
$r=1$ $$\P_s\eta_1=\P_s\X_1=
\sum^{\infty}_{j=1} C^j_s \P^j\X_{s+1-j}+
\sum^s_{k=2}B^k_s\sum^{s-k}_{j=0} C^j_{s-k}\P^j\X_{1+s-j-k}=$$
$$\sum^s_{j=1} C^j_s\P^j\X_{s+1-j}-
\sum^{\infty}_{k=2}\Bigl(\sum^\infty_{n=1}\sum_{i_1+\dotsb +j_n=k} P_s
{\pmatrix i_1&\ldots & i_n\\j_1 & \ldots & j_n\endpmatrix}
\P^{j_1}\eta_{i_1}\dotsb \P^{j_n}\eta_{i_n}\Bigr)\cdot$$
$$\Bigl(\sum_{j=0}^{s-k}C^j_{s-k}\P^j\X_{1+s-j-k}\Bigr)=
\sum^\infty_{n=1}\sum P_s \pmatrix i_1 &\ldots & i_n\\
j_1 &\ldots & j_n\endpmatrix
\P^{j_1}\eta_{i_1}\dotsb \P^{j_n}\eta_{i_n},$$
where the second sum in the last formula is taken by all matrices
$\pmatrix i_1 &\ldots & i_n\\j_1 &\ldots & j_n\endpmatrix$ such that
$i_1+\dotsb +i_n+j_1+\dotsb +
j_n=s+1, \ i_m\geqslant 1, \ j_m\geqslant 0$.
According to lemma 1 in this sum it is sufficient consider
only matrices, where $j_m>0$ for all $m$.

Prove now the lemma for $r=N$, considering that it is proved for
$r<N$. According to (3)
$$\P_s\Bigl(\sum^\infty_{n=1}{\frac {1}{n!}}\sum_{i_1+\dotsb+i_n=r}
\eta_{i_1}\dotsb \eta_{i_n}\Bigr)=
\sum^{s+r-1}_{j=1}C^j_{s}\P^j\X_{s+r-j}+
\sum^\infty_{k=2}B^k_s\sum_{j=0}^{s-k} C^j_{s-k}\P^j\X_{r+s-j-k}.$$
Thus according to lemma 3, lemma 1 and inductive hypothesis,
$$\P_s\eta_r=\sum^{\infty}_{j=1}C^j_s\P^j
\Bigl(\sum^\infty_{n=1}{1\over n!}\sum_{i_1+\dotsb +i_n=s+r-j}
\eta_{i_1}\dotsb \eta_{i_n}\Bigr)+$$
$$+\sum^{\infty}_{k=2}\Bigl(\sum_{i_1+\dotsb +j_n=k} P_s
{\pmatrix i_1 &\ldots & i_n\\j_1 &\ldots & j_n\endpmatrix}
\P^{j_1}\eta_{i_1}\dotsb \P^{j_n}\eta_{i_n}\Bigr)\cdot$$
$$\sum_{j=0}^{s-k}C^j_{s-k}\P^j
\Bigl(\sum^\infty_{n=1}\frac {1}{n!}\sum_{i_1+\dotsb+i_n=r+s-j-k}
\eta_{i_1}\dotsb \eta_{i_n}\Bigr)
-\P_s\Bigl(\sum_{n=2}^{\infty}\frac{1}{n!}\sum_{i_1+\dotsb+i_n=r}
\eta_{i_1}\dotsb \eta_{i_n}\Bigr)=$$
$$=\sum^\infty_{n=1}\sum P_s
{\pmatrix i_1&\ldots & i_n\\j_1 &\ldots & j_n\endpmatrix}
\P^{j_1}\eta_{i_1}\dotsb \P^{j_n}\eta_{i_n},$$
where the second sum in the last formula is taken by all matrices
$\pmatrix i_1 &\ldots & i_n\\ j_1 &\ldots & j_n\endpmatrix$ such that
$i_1+\dotsb +i_n+j_1+\dotsb +
j_n=s+r$, $i_m\geqslant 1, j_m\geqslant 1$. $\square$

\vskip 0.4cm
\noindent
\centerline{5. KP hierarchy}
\vskip 0.4cm
\noindent

According to [DKJM] the Bacher-Akhiezer function $\F$ is
$$\F(x,k)=\text{exp}(\sum x_j k^j)\frac{\tau(x_1-k^{-1},
x_2-\frac 12 k^{-2}, x_3-\frac 13 k^{-3},...)}{\tau(x_1,
x_2, x_3,...)}$$ for some function $\tau(x_1, x_2,...)$.
By analogy of [N] this gives a possibility to
describe the KP hierarchy as an infinite system of
differential equations on $v(x,k)=-\ln \tau(x,k)$.
Really
$$\sum^\infty_{j=1}\eta_j k^{-j}=\ln\F(x,k)-\sum^\infty_{j=1}
x_j k^j=-v(x_1-k^{-1}, x_2-\frac12 k^{-2},...)+v(x)=$$
$$\sum^\infty_{n=1}\sum_{i_1+\dotsb+i_n=j}\frac{(-1)^{n+1}}
{n!i_1\dotsc i_n}\P_{i_1}\dotsb \P_{i_n} v(x) k^{-j}.$$
Therefore
$$\eta_r=\sum^\infty_{n=1}\sum_{i_1+\dotsb+i_n=r}\frac{(-1)^{n+1}}
{n!i_1\dotsc i_n}\P_{i_1}\dotsb \P_{i_n} v. \tag 4$$

\vskip 0.4cm
\noindent
{\bf Theorem 2.} {\sl There exist universal rational
coefficients
$$R_r\pmatrix s_1 &\ldots & s_n\\ t_1 &\ldots & t_n\endpmatrix,
R_{ij}\pmatrix s_1 &\ldots & s_n\\t_1 &\ldots & t_n\endpmatrix$$
such that
$$\eta_r={1\over r}\P_rv+\sum^\infty_{n=1}\sum
R_r\pmatrix s_1 &\ldots & s_n\\t_1 &\ldots & t_n\endpmatrix
\P_{s_1}\P^{t_1}v\dotsb \P_{s_n}\P^{t_n}v , \tag 5$$
$$\P_i\P_j v=\sum^\infty_{n=1}
\sum R_{ij}\pmatrix s_1 &\ldots & s_n\\t_1 &\ldots & t_n\endpmatrix
\P_{s_1}\P^{t_1}v\dotsb \P_{s_n}\P^{t_n}v, \tag 6$$
where the second sums are taken by all matrices
$\pmatrix s_1 &\ldots & s_n\\ t_1 &\ldots & t_n\endpmatrix$ such that
$s_m, t_m\geqslant 1$, and the sum $s_1+\dotsb +s_n+t_1+
\dotsb +t_n$ is equal $r$ for (5) and $i+j$ for (6).
Moreover $$R_{ij}\pmatrix s_1 &\ldots & s_n\\ 1 &\ldots &
1\endpmatrix= \frac{j}{s_1\dotsb s_n} P_i(s_1...s_n).$$}

\vskip 0.4cm
\noindent
Proof: An induction by $k$  and $i+j$. For $i+j=2$ the theorem
is obviously. For $r=1$ it follows from (4).
Prove the theorem for $i+j=N$ and $r=N-1$, considering that it is
proved for $i+j<N$ and $r<N-1$.
Later we consider that $s_m, t_m\geqslant 1$ and
$\SI_n=s_1+\dotsb +s_n+t_1+\dotsb +t_n$.
Then according to (4) and (6)
$$\eta_r={1\over r}\P_rv+\sum^\infty_{n=2}\sum_{s_1+\dotsb+s_n=r}
\frac{(-1)^{n+1}}{n! s_1\dotsb s_n}\P_{s_1}\dotsb \P_{s_n}v(x)=$$
$${1\over r}\P_rv+\sum^\infty_{n=1}\sum_{\SI_n=r}
R_r\pmatrix s_1 &\ldots & s_n\\t_1 &\ldots & t_n\endpmatrix
\P_{s_1}\P^{t_1}v\dotsb \P_{s_n}\P^{t_n}v.$$
Thus according to (5), (6) and lemma 4
$${1\over j}\P_i\P_j v=\P_i\eta_j-\P_i
\Bigl(\sum^\infty_{n=1}\sum_{\SI_n=j}
R_j\pmatrix s_1 &\ldots & s_n\\t_1 &\ldots & t_n\endpmatrix
\P_{s_1}\P^{t_1}v\dotsb \P_{s_n}\P^{t_n}v\Bigr)=$$
$$\sum^\infty_{n=1}\sum_{\SI_n=i+j}
P_i\pmatrix s_1 &\ldots & s_n\\t_1 &\ldots & t_n\endpmatrix
\P^{t_1}\eta_{s_1}\dotsb \P^{t_n}\eta_{s_n}-$$
$$\P_i\Bigl(\sum^\infty_{n=1}\sum_{\SI_n=j}
R_j\pmatrix s_1 &\ldots &s_n\\t_1 &\ldots & t_n\endpmatrix
\P^{t_1}\P_{s_1}v\dotsb \P^{t_n}\P_{s_n}v\Bigr)=$$
$$\sum^\infty_{n=1}\sum_{s_1+\dotsb+ s_n+n=i+j}
P_i\pmatrix s_1 &\ldots & s_n\\1 &\ldots & 1\endpmatrix
\P(\frac{1}{s_1}\P_{s_1}v)\dotsb \P(\frac{1}{s_n}\P_{s_n}v)+$$
$$\sum^\infty_{n=1}\sum_{\SI_n=i+j,t_1+\dotsb+ t_n>n}
R_{ij}\pmatrix s_1 &\ldots & s_n\\t_1 &\ldots & t_n\endpmatrix
\P_{s_1}\P^{t_1}v\dotsb \P_{s_n}\P^{t_n}v.$$

The obvious relation $P_{i}\pmatrix s_1 &\ldots & s_n\\1 &\ldots & 1\endpmatrix=
P_i(s_1...s_n)$ conclude the proof. $\square$

\vskip 0.4cm
\noindent
{\bf Remark.}  The first equation from (6) is the
"integrated over $x_1$" KP equation
$$\P^2_2 v=\frac 43\P_3\P_1 v-\frac 13\P^4_1 v+2(\P^2_1 v)^2.$$
The relation (6) was first deduced in [ DN ] from the Hirota
hierarchy [ DKJM ]. This method gives another formulas for
coefficients $R_{ij}$. The equalities between the
coefficients provide nontrivial combinatorial identities.
$\square$

The system (6) is called the {\sl KP hierarchy}.
{\sl The dispersionless KP hierarchy}
is  called  the  infinite  system  of  differential equations
which appears from (6) if to throw out all monomials
$$R_{ij}\pmatrix s_1 &\ldots & s_n\\t_1 &\ldots & t_n\endpmatrix
\P_{s_1}\P^{t_1}v\dotsb \P_{s_n}\P^{t_n}v,$$ where
$\sum^n_{i=1}t_j>n$. The theorem 2 gives

\vskip 0.4cm
\noindent
{\bf Consequence} {\sl The despersionless KP hierarchy is
$$\frac 1j\P_i\P_j v=\sum^\infty_{m=1}\sum_{s_1+\dotsb +s_m=
i+j-m} P_i(s_1...s_m)\frac{1}{s_1}\P\P_{s_1}v\dotsb\frac {1}{s_m}
\P\P_{s_m}v. \tag 7$$}

\vskip 0.4cm
\noindent
\centerline{6. Proof of theorem 1}
\vskip 0.4cm
\noindent

According to [DVV, K, W2] one of $A_{n-1}$-potentials is
$$\W F(x_1,...,x_{n-1})=v(x_1,\frac{x_2}{2},...,\frac{x_{n-1}}
{n-1}, 0,...),$$ where $v$ satisfies the
dispersionless KP hierarchy and $\P_nv=0$ (dispersionless
Gelfand-Dikii equation).

According to (7) from this follows
$$0=\P_\ell\P_n \W F=\sum^\infty_{m=1}
\sum_{s_1+\dotsb + s_m=\ell+n-m} P_n(s_1\dots s_m)
\P\P_{s_1}\W F\dotsb\P\P_{s_m}\W F=$$
$$n\P\P_{n+\ell-1} \W F +\sum^\infty_{m=2}
\sum_{s_1+\dotsb + s_m=\ell+n-m} P_n(s_1\dots s_m)
\P\P_{s_1}\W F\dotsb\P\P_{s_m}\W F.$$
Thus $$\P\P_{n+\ell-1} \W F=-\frac{1}{n}\sum^\infty_{m=2}
\sum_{s_1+\dotsb + s_m=\ell+n-m} P_n(s_1\dots s_m)
\P\P_{s_1}\W F\dotsb\P\P_{s_m}\W F.\tag 8$$
According to [W2]
$$\P\P_s\W F=x_{n-s}\quad \text{and}\quad \P\P_{n+s}
\widetilde F=-s\P_s\widetilde F$$ for $s<n$. In additional
$\P\P_n\W F=0$ and thus according to (8)
$\P\P_{n+\ell-1}\W F=x_{1-\ell}$ for all $\ell$.
Therefore $$\P_s\widetilde F=-\frac 1s x_{-s}=\P_sF_A$$ and
$F_A - \W F$ --- is a linear form.
According to [D2, p.16] this form is equal 0.
According [Z] $F_B(x_1,...,x_n)=F_A(x_1, 0, x_2,
0,...,x_{n-1}, 0, x_n). \square$

There are the potentials finding by our formulas
$$(A_2)\quad F=\frac 12 x^2_1 x_2+\frac{1}{24} x_2^4,$$
$$(A_3)\quad F=\frac 12 x^2_1 x_3+\frac{1}{2} x_1 x_2^2+
\frac 14 x^2_2 x_3^2+\frac{1}{60} x_3^5,$$
$$(A_4)\quad F=\frac 12 x^2_1 x_4+ x_1 x_2 x_3+
\frac 16 x^3_2+\frac{1}{12} x_3^4+\frac{1}{2} x_2 x_3^2 x_4+
\frac{1}{4} x_2^2 x_4^2+\frac{1}{6} x_3^2 x_4^3+\frac{1}{120} x_4^6,$$
$$(A_5)\quad F=\frac 12 x^2_1 x_5+\frac 12 x^2_2x_3+
\frac 14 x_2^2 x^2_5+x_2 x_3 x_4 x_5+x_1 x_2 x_4+
\frac 16 x_2 x^3_4+\frac 12 x_1x_3^2+\frac 12 x^2_3 x_4^2+$$
$$+\frac 16 x_3^3 x_5+\frac 16 x_3^2 x_5^3+\frac 12 x_3 x_4^2 x^2_5+
\frac 16 x_4^4 x_5+\frac 18 x_4^2 x_5^4+\frac{1}{210} x^7_5.$$
$$(B_2)\quad F=\frac 12 x^2_1 x_2+\frac{1}{60} x_2^5,$$
$$(B_3)\quad F=\frac 12 x^2_1 x_3+\frac{1}{2} x_1 x_2^2+
\frac 16 x^3_2 x_3+\frac{1}{6}x^2_2 x_3^3+\frac{1}{210} x^7_3.$$

\vskip 0.4cm
\centerline{References}
\vskip 0.4cm

[D1]  B.Dubrovin, Geometry of 2D topological field
theories, In: "Integrable Systems ans Quantum Groups",
Eds. M.Francaviglia, S.Greco,Springer Lecture Notes in Math. 
1620 (1996), 120-348.

[D2]  B.Dubrovin, Painleve transcendents in two-dimensional
topological field theory. Pre\-print SISSA 24/98/FM P.107.

[DKJM]  E.Date, M.Kashiwara, M.Jimbo, T.Miwa, Transformation groups 
for soliton equa\-tion, Proceedings
of RIMS Symposium on Non-Linear Integrable System.
Singapore: World Science Publ. Co., 1983, 39-119.

[DN]  B.A.Dubrovin, S.M.Natanzon, Real theta-function solutions
of the Kadomtsev-Petvi\-ashvili equation. Math. USSR Irvestiya,
32:2 (1989), 269-288.

[DVV]  R.Dijkgraaf, E.Verlinde, H.Verlinde, Topological strings
in $d<1$. Nucl. Phys., B 352 (1991), 59.

[K]  I.Krichever, The dispersionless Lax equations and topological
minimal models. Com\-mun. Math. Phys. 143 (1992), 415-429.

[N]  S.M.Natanzon, Differentil equations on the Prym theta function.
A realness criterion for two-dimensional, finite-zone, potential
Schr\"odinger operators. Functional Anal. Appl. 26:1 (1992), 13-20.

[W1]  E.Witten, On the structure of the topological phase of 
two--dimensional gravity. Nucl. Phys., B340 (1990), 281-332.

[W2]  E.Witten, Algebraic geometry associated with matrix models
of two dimensional gravity, Topological models in modern mathematics
(Stony Brook, NY, 1991), Publish or Perish, Houston, TX 1993, 235-269.

[Z]  J.-B.Zuber, On Dubrovin's topological field theories.
Mod. Phys. Lett. A9 (1994) 749-760.

Moscow State University, Moscow, Russia.

Independent University of Moscow, Moscow, Russia.

e-mail:   natanzon\@mccme.ru

\end